\documentclass[twocolumn,aps,ajp,groupedaddress]{revtex4}
\usepackage[pdftex]{graphicx}
\usepackage{graphics}
\usepackage{amsmath,amsthm,amsfonts,amssymb}
\usepackage{longtable}

\begin{document}
\title{On the experimental determination of the one-way speed of light}
\author{Israel P\'erez }
\email{cooguion@yahoo.com}
\affiliation{Centro de Investigaci\'on Cient\'ifica y de Educaci\'on Superior de Ensenada,\\ Carretera Ensenada-Tijuana 3918, Zona Playitas, C.P. 22860 Ensenada B.C. M\'exico.\\ Optics Department, Applied Physics Division}

\begin{abstract}
In this contribution the question of the isotropy of the one-way speed of light from an experimental perspective is addressed. In particular, we analyze two experimental methods commonly used in its determination. The analysis is aimed at clarifying the view that the one-way speed of light cannot be determined by techniques in which physical entities close paths. The procedure employed here will provide epistemological tools such that physicists understand that a direct measurement of the speed not only of light but of any physical entity is by no means trivial. Our results shed light on the physics behind the experiments which may be of interest for both physicists with an elemental knowledge in special relativity and philosophers of science.
\end{abstract}
\maketitle

\section{Introduction}
Before the XVII century most people used to believe that the speed of light $c$ was infinite. This belief, however, started to change when Ole R\"oemer did the first estimation by means of the observation of the eclipses of Jupiter's satellite Io. Still some physicists argue \cite{greavesx,goldstein} that this method provides a direct determination of the isotropy of light in one direction. Nevertheless, Karlov \cite{karlov} has given strong arguments to show that R\"oemer's approach, in fact, constitutes a two-way measurement of the speed of light. Supporting this position Zhang \cite{zhang} has theoretically discussed that not any conceivable experiment can succeed in measuring the one-way speed of light and more recent articles \cite{sabat,guerra1,guerra2,abreu,spavieri,iyer1} point in this same direction. 

Since the pioneers works of Michelson and Morley many reports \cite{michelson1,page,bergel,essen,cedarholm,babcock,jaseja,waddoups,brillet,herrmann,bates,huang,bay,evenson3,rowley,muller,muller1,wolf,antonini,hollberg} have claimed the testing of the second postulate of special relativity (SR) \cite{einstein2}. However, all of  these cases have a common factor: \emph{that light follows closed paths} and therefore they only test, directly or indirectly, the isotropy of the average round trip speed or the so called \emph{two-way speed of light}. So far, few authors \cite{riis,krisher,will,feenberg,fung} have propounded very clever thought experiments to test the isotropy of the one-way speed of light but not general consensus has been achieved. In a recent paper E. Greaves et al. \cite{greaves} reported the achievement of a measurement of the one-way speed of light. A claim that not only has raised severe critics \cite{klauber,finkelstein} but also certainly goes against the opinion of the vast majority of specialists \cite{zhang,sabat,guerra1,guerra2,abreu,spavieri,iyer1,perez1,grunbaum,reichenbach,townsend,ungar}. 

From our perspective it is worthy to discuss in great detail not only the experimental methods but the arguments that lead researchers to conclude that the measurement of the one-way speed of light is feasible. In this article we meticulously analyze two experimental methods commonly used in physics laboratories in the measurement of the speed of light. From this analysis representative expressions of the problem will be derived for the one-way and two-way speed of any physical entity (PE). By doing this we shall endeavor to show that such methods are actually of the two-way type since the PEs involved close paths and, therefore, claims on the measurement of the one-way speed of light founded on these methods lose their validity. 

Even nowadays some teachers and scientists still believe that the determination of the speed of light reduces to specifying a distance $s$ and reckoning the speed as $c=s/t$, where $t$ the time taken by light during its journey. However, when the problem involves two inertial systems of reference, our investigation shows that the measurement of the one-way speed of light is by no means as trivial as at first sight appears. This analysis will help us to understand why the measurement of the one-way speed of light has been elusive. 

The insight developed here may be easily implemented not only in an introductory course of SR but also in any physics laboratory widening the view of SR presented in textbooks and scientific papers. And, at the same time, it may constitute a point of departure for the invention of more effective methodologies in the experimental determination of the one-way speed of light. Lastly, although the work is intended for experimental physicists and philosophers of science, anyone with a basic course in SR can easily grasp its contents.

\section{Preliminaries}
\label{preli}
\subsection{One-way speeds and measured speeds}
One of the aims of the present investigation is to analyze the measuring processes and, consequently verify whether the experimental techniques allow us to know the one-way speed of the PEs. For convenience, we shall denote the ``\emph{measured speeds}" with a bar above the quantity, e.g. $\bar{v}$. The reason for this is just to make a clear epistemological distinction between these quantities. The features that distinguish the one-way speeds from the measured speeds will be logically assimilated as we advance. Furthermore, we shall restrict ourselves to study only \emph{direct} measurements of speed, i.e., by measuring space and time. Indirect measurements of speed like, for instance, $v=p/m$, where $p$ is the momentum and $m$ the mass of the particle, are not considered here. 

\subsection{A matter of semantics}

We should warn the reader that the subject may prove somewhat difficult to understand if we do not make clear the following peculiarity. SR is based on two postulates, namely: (1) the principle of relativity and (2) the constancy of the ``\emph{speed of light}" for all inertial systems of reference. Usually, the Lorentz transformations are derived from these postulates, however, one can derive them considering only the first postulate so long as one adopts an adequate clock synchronization for the inertial systems \cite{schroder,mermin,rindler}. When one proceeds in this way there remains a universal constant $c$, with the dimensions of velocity and of finite value, to be determined by experiment. Analogously, Maxwell's electrodynamics (ME) has a constant $V=1/\sqrt{\mu_0\epsilon_0}$ representing the speed of electromagnetic waves (EMW) in vacuum to be determined also by experiment. It is worth noting that neither SR nor ME are capable of determining the value of their respective constants without relation to experiment. However, by convention, the Comit\'e International des Poids et Mesures (BIPM) \cite{giacomo0} defined the value of the speed of EMW  as $V \equiv 299\,792\,458$ m/s. But this does not imply that the actual speed of EMW possesses that exact value but that the actual value is around $V$ with a speed uncertainty within the interval $0<a<1$ m/s, with $a\in \Re$. Indeed, the actual speed of EMW seems to be nearly a constant (within the limits of experimental accuracy) but only when measured in the radiation zone \cite{jackson}, whilst in the near and intermediate zones the speed of EMW may acquire other values even in vacuum \cite{mugnai,budko}. And by convention, again, $c$ was identified with the constant $V$ of ME. But why did SR borrow the value from another theory (ME)? Why is not SR capable of determining the value of its own constants? This question can be also asked to other theories like, for instance, the general theory of relativity but this kind of problems cannot be treated here, the answer can be found elsewhere \cite{ellis,mendelson,narlikar}. What is important to make clear here is that to avoid semantical misunderstandings we shall consider that $c=V$ and, as we shall see later, the definition of the constant $c$ can affect the outcome of a measurement.

In summary, based on ME, we make the assumptions that at least there exists one inertial system of reference $K$ where the one-way speed of light in vacuum is isotropic and maximal in the radiation zone and such vacuum is isotropic and homogenous. Following the jargon of Iyer and Prabhu \cite{iyer1}, this system shall be called the \emph{isotropic system}. Nevertheless, we shall explain that the experimental methods to be treated here do not allow us to know the one-way speed of any PE for inertial systems in motion relative to $K$. 

\section{Methodologies for the determination of the speed of any physical entity}
\label{measmo}

To appreciate the importance on how the experimental techniques influence the outcomes of a measurement, we shall analyze two of the most common methodologies adopted for the determination of the speed of any PE, being the speed of light just one particular case. And without loss of generality the same principles propounded here can be applied for the analysis of any other experimental method. The methods will be labeled as 1 and 2, respectively. 

\subsection{Method 1}
This method resembles the one carried out by Aoki et al. \cite{aoki} in the determination of the speed of light, however, as we shall show below, their method only tells us only one part of the whole result.

\subsubsection{A non-trivial measurement}
\begin{figure}[hbp]
\begin{center}
\includegraphics[width=3.5cm]{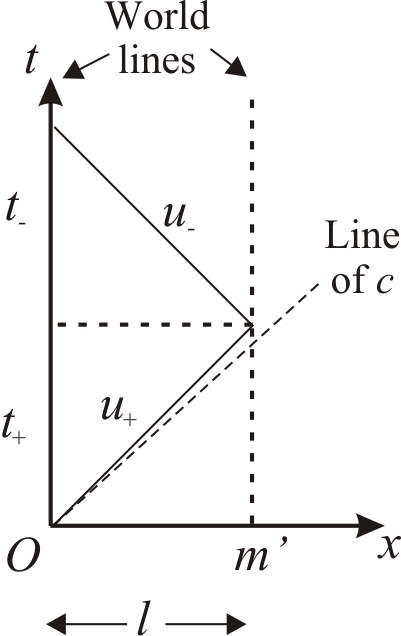}
\caption{Space-time diagram for the measurement of speed in the isotropic frame. The observer situated at the origin has to wait for the returning information that travels at the speed $u_-$ and thus he measures the total time $t=t_{+}+t_{-}$.}
\label{isotro}
\end{center}
\end{figure}

First of all, it is assumed that the measurement is carried out in the inertial system $K$. Let us then consider that an observer, who has placed a clock at the origin $O$, is interested in measuring the one-way speed $u_{+}$ of a PE$_{+}$ by measuring the time it takes to travel an arbitrary distance $l$ that extends from $O$ to any desire point $m'$ (see Figure \ref{isotro}). 
But now here we must ask: How would the observer know that the PE$_{+}$ has arrived at the opposite endpoint? Certainly, since the clock is at $O$ the information of the arrival event has to return to $O$ by any physical means at any one-way speed $u_{-}$. This can be achieved, for instance, by just observing the event, that is, by means of a light signal, or perhaps by putting a mirror or a receiver (detector) or any other instrument that senses the event of arrival and in turn sends a returning signal towards the origin. In any case, \emph{the clock ought to measure the round trip time}. Imagine that at $t_0=0$ we let the PE$_{+}$ in question to depart from $O$ and traverse the distance $l$ along the $x$-axis. At the opposite endpoint we place a contrivance at $m'$ that receive the PE$_{+}$ and returns the arrival information via any other PE$_{-}$ (it could be the same PE) towards the origin, where the observer, measures the time $t$ that it takes to complete the round trip \footnote{For purposes of illustration, we assume the response time of the receiver is negligible in comparison to the times spent in the journeys by the PEs.}. Bearing this in mind, the outward time or \emph{time of flight} of the PE$_{+}$ is $t_{+}=l/u_{+}$, and the time of the returning information or \emph{delay time} is $t_{-}=l/u_{-}$. Hence the one-way measured speed for the PE$_{+}$ is simply
\begin{equation}
\label{velk1}
\bar{u}_{+}=l/(l/u_{+})=u_{+}.
\end{equation}
And for the returning information the one-way measured speed is
\begin{equation}
\label{velk2}
\bar{u}_{-}=l/(l/u_{-})=u_{-}.
\end{equation}
It is clear that the measured speeds are equal to the one-way speeds. But the measurement is done only when the information returns to $O$, hence, $t=t_{+}+t_{-}$ and the measured speed is 
\begin{equation}
\label{uaver}
\bar{u}=\frac{2l}{l/u_{+}+l/u_{-}}=\frac{2}{1/u_{+}+1/u_{-}}=\frac{2u_{+}u_{-}}{u_{+}+u_{-}}.
\end{equation}
Note that this expression is the harmonic mean of the speed and if $u_+\neq u_-$ the measured speed does not correspond to the one-way speed of the PE$_+$. But if $u_+=u_-=u\le c$ then $\bar{u}=u$. In the case of Aoki et al. \cite{aoki} in which they used light in both directions one may assume that $u_+= u_-= c$ thus one expects that $\bar{u}= c$ which corresponds to the one-way speed of light. At first sight the previous operation appears to be trivial but this is not the case. Next we show why.

\begin{figure}[htp]
\begin{center}
\includegraphics[width=5cm]{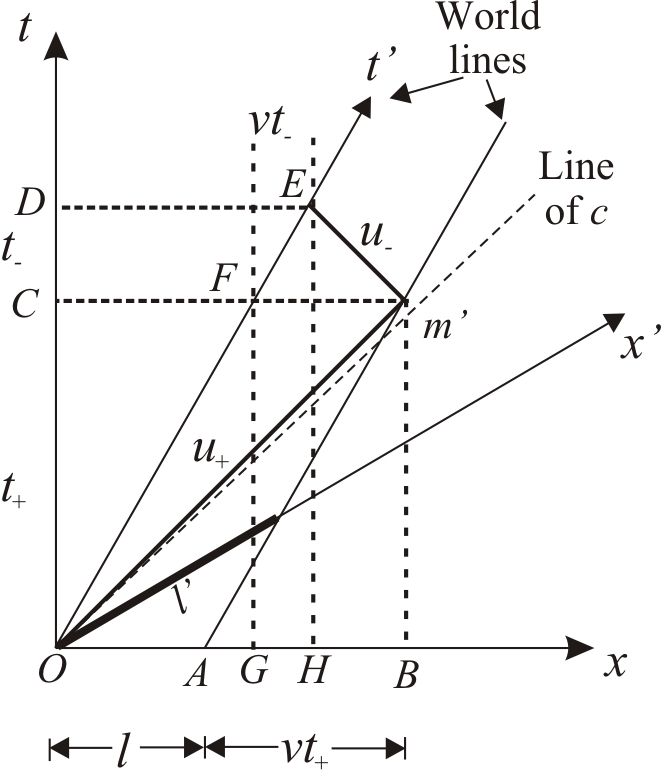}
\caption{Space-time diagram for the measurement of the speed realized in the frame $K'$ as seen by the isotropic frame.}
\label{stmeas}
\end{center}
\end{figure}
\subsubsection{Longitudinal motion}

Now let us imagine that an observer in an inertial system $K'$, which is moving along the $x$-axis relative to the system $K$ at speed $v<c$, wants to determine with the same experimental setup not only the speed of the PE$_+$ but also the speed of $K'$ relative to $K$. Figure \ref{stmeas} shows the space-time diagram for this problem. For simplicity we shall assume that the measurement starts at $t_0=t'_0=0$. As judged from $K$, the PE$_{+}$ follows the path $Om'$, whereas the PE$_{-}$ follows the path $m'E$. From the figure we have that $u_+=OB/OC=OB/t_{+}$ and $u_-=HB/CD=HB/t_{-}$. Also $v=AB/OC=AB/t_{+}=GH/CD=GH/t_{-}$ and $OA=GB=l$. From these expressions we can derive the following relations
\begin{eqnarray}
\label{rmas2}  OB & = & u_+t_{+}=OA+AB=l+vt_{+}, \nonumber \\
\label{rmenos1} HB & = & u_-t_{-}=GB-GH=l-vt_{-}. \nonumber
\end{eqnarray}
Solving for the times we have that $t_{\pm}=l/(u_{\pm}\mp v)$. To determine the times spend in each journey, as judged in $K'$, we just have to consider length contraction $l=l'\gamma^{-1}$ and time dilation $t=\gamma t'$, where $\gamma=1/\sqrt{1-v^2/c^2}$. It follows that the observer in $K'$ determines that
\begin{equation}
\label{tonpar}
t'_{\pm}=\gamma^{-2}\frac{l'}{u_{\pm}\mp v}.
\end{equation}
Hence the one-way measured speeds are
\begin{equation}
\label{conpar}
\bar{u}'_{\pm}=\frac{l'}{t'_{\pm}}=\gamma^{2}(u_{\pm}\mp v).
\end{equation}
By comparison with the case in $K$ the one-way measured speed does not correspond to the one-way speed $u_{\pm}$. 
But since the observer in $K'$ can only measure the round trip time $t'=t'_{+}+t'_{-}$, he will, in fact, measure $\bar{u}'=2l'/t'$ or explicitly
\begin{eqnarray}
\label{tottim}
\bar{u}'&= &\frac{2l'}{\gamma^{-2}\frac{l'}{u_{+}-v}+\gamma^{-2}\frac{l'}{u_{-}+v}} \nonumber\\
&=&\frac{2}{1/\bar{u}'_{+}+1/\bar{u}'_{-}}  =  \frac{2\gamma^{2}(u_{+}-v)(u_{-}+v)}{u_{+}+u_{-}}.
\end{eqnarray}
In any case if $v=0$ the expressions reduce to the case at rest.  Note further that the measured speed not only corresponds to the average round trip speed (the harmonic mean of the speed) which in turn is function of $v$ but also depends on the definition of $c$. Hence \textbf{this experimental procedure does not allow the observer in $K'$ to determine by himself neither the one-way speeds $u_{\pm}$ nor $v$}.
\begin{figure}[htp]
\begin{center}
\includegraphics[width=5cm]{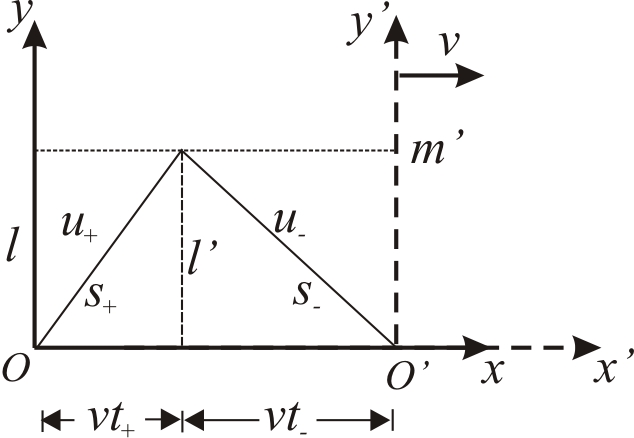}
\caption{Measurement conducted in $K'$ for the transversal motion as seen from the isotropic frame. Only the spatial dimensions are considered.}
\label{transmot}
\end{center}
\end{figure}
\subsubsection{Transversal Motion}

Now let us further imagine that the experimental arrangement has been rotated $\pi/2$ rad and we want to perform the same measurment. If the apparatus is placed in the isotropic system this operation is trivial and it reduces to Eqs. \eqref{velk1}-\eqref{uaver}. But when the experiment is conducted in $K'$ it acquires a distinct aspect. The spatial situation is depicted in Fig. \ref{transmot}. As seen from $K$, the PE$_{+}$, with speed $u_+$, arrives at the point $m'$ from where the PE$_{-}$, with speed $u_-$, is sent towards the origin $O'$. To determine the time for each journey, we use the Pythagorean theorem for the distances $S_{\pm}$ traveled by the entities during each journey. Hence, we have that $S_{\pm}^2=(u_{\pm}t_{\bot \pm})^2=(vt_{\bot \pm})^2+l^2$. On solving for the transversal times we obtain $t_{\bot \pm}=(l/u_{\pm})\gamma_{\pm}$, where $\gamma_{\pm}=1/\sqrt{1-v^2/u_{ \pm}^2}$. Note also that 
\begin{equation}
\label{wwe}
u_{\pm}=\sqrt{u_{x\pm}^2+u_{y\pm}^2}, \quad u_{x\pm}=v;  \quad u_{y\pm}=u_{\pm}\gamma^{-1}_{\pm}.
\end{equation}
Since the length $l$ is perpendicular to the line of motion it follows from relativistic effects that $l=l'$ and $t_{\bot}=t'_{\bot}\gamma$, hence
\begin{equation}
\label{otrt}
t'_{\bot \pm}=\frac{l'}{u_{ \pm}}\gamma_{\pm}\gamma^{-1}.
\end{equation}
Consequently, the one-way measured speeds in the transversal direction are
\begin{equation}
\label{uss}
\bar{u}'_{\bot \pm}=u_{ \pm}\gamma^{-1}_{\pm}\gamma=\gamma u_{y\pm}.
\end{equation}
But he can only measure the round trip time $t'_{\bot}$ and the measured speed is given by
\begin{equation}
\label{twotrans}
\bar{u}'_{\bot}=\frac{2l'}{t'_{\bot}}=\frac{2}{1/\bar{u}'_{\bot+}+1/\bar{u}'_{\bot-}}=\frac{2\gamma u_+u_-}{\gamma_+u_-+\gamma_-u_+}.
\end{equation}
Once again this is the harmonic mean of the speed in the transversal direction and when $v=0$ the expressions reduce to the case at rest. Also the observer in $K'$ cannot solve for the one-way speeds $u_{\pm}$ and $v$. Let us see whether this trouble can be resolved.

\subsubsection{Determination of the speeds}
\label{truespe}
So far the observer in $K'$ still has three unknowns to find, namely: $u_{\pm}$, $v$. With the aid of Equations \eqref{tottim} and \eqref{twotrans}, however, two speeds can be estimated if we constrain the experimental situation to the following conditions. (1) Both measurements, longitudinal and transversal, are carried out ``simultaneously", therefore, now we need a new setup with four paths, i.e., two for the forward journeys and two for the backward ones. (2) The medium for the displacement of the four PEs is isotropic and homogeneous and its temperature remains constant. (3) The previous point helps to  guarantee that, if we use the same PEs for the four paths, the speed of the PEs must be approximately the same, hence we might assume that $u\equiv  u_{+}\approx u_{-}$. For instance, the physical entities could be electric fields traveling through ``identical wires" or light signals in vacuum. If these conditions are satisfied expression \eqref{tottim} becomes
\begin{equation}
\label{asdf}
\bar{u}'=u\gamma^{2}\gamma^{-2}_u,
\end{equation}
whereas expression \eqref{twotrans} reduces to  
\begin{equation}
\label{twotra}
\bar{u}'_{\bot}=u\gamma\gamma^{-1}_u.
\end{equation}
From the previous expressions we can solve for $u$ and $v$ in terms of the measured quantities, that is,
\begin{equation}
\label{sdfa}
u=\frac{\bar{u}_{\bot}'^2}{\bar{u}'}
\end{equation}
and
\begin{equation}
\label{adfe}
v=\frac{c\bar{u}_{\bot}'}{\bar{u}'}\sqrt{\frac{\bar{u}_{\bot}'^2-\bar{u}'^2}{c^2-\bar{u}_{\bot}'^2}}.
\end{equation}
It is to be noted that the value of $v$ depends on the definition of $c$. If we imagine that $u= c$ then we would expect that $\bar{u}'=\bar{u}'_{\bot}=\bar{u}'_{\bot \pm}=c$. This result justifies the constancy of the two-way speed of light for this method. However, the one-way measured speeds $\bar{u}'_{\pm}$ remain velocity dependent. For this reason we shall call the system $K'$ the \emph{anisotropic system}. If we follow this line of thought, then any inertial system in motion relative to the isotropic system is also an anisotropic system.

On the other hand, our expressions are in terms of velocities but they can also be put in terms of the times. Thus, it is not difficult to foresee that this setup can be easily reproduced in the laboratory. For instance, we may use a four-channel oscilloscope connecting four cables two of them along the $x'$-axis and the other two along the $y'$-axis forming a right angle, so we can determine $u$ and $v$. This procedure allows us to use light signals instead of cables. Aoki et al. \cite{aoki} applied this method and the value they reported only corresponds to one orientation of the apparatus (say, longitudinal). The case in two orientations would resemble a Michelson-Morley experiment.
\begin{figure}[htp]
\begin{center}
\includegraphics[width=3.7cm]{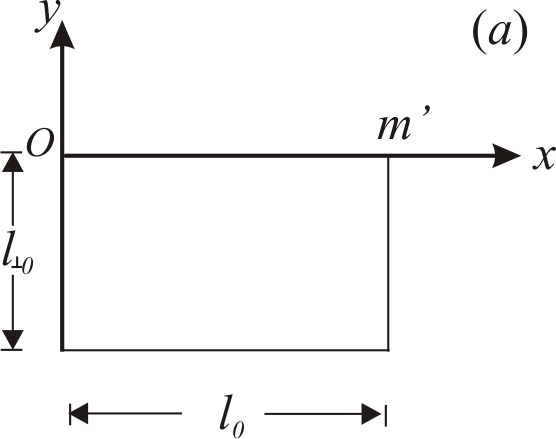}\hspace{0.5cm} \includegraphics[width=4.2cm]{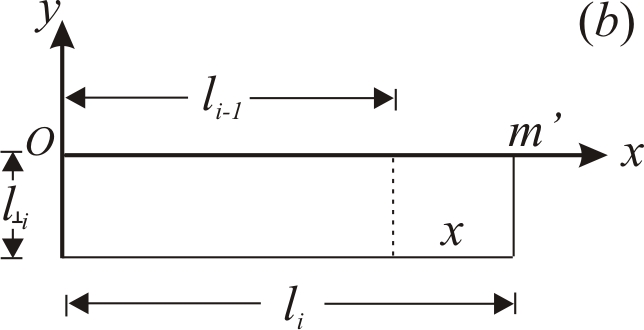}
\caption{Measurement conducted in $K$. (a) Initial setup in a rectangular geometry with $L=l_0+2l_{\bot 0}$. (b) The detector has been displaced to the position $l_i$ and the rectangle has been modified with $L=l_i+2l_{\bot i}$.}
\label{fig4_iperez}
\end{center}
\end{figure}
\subsection{Method 2}

\subsubsection{Measurement at rest in $K$}
\label{metho2}
Now let us consider the other method that was used by Greaves, Mugnai, Budko et al. \cite{greaves,mugnai,budko}. First assume that we are in the isotropic system $K$. For simplicity in our analysis we shall assume that the paths to be followed by the PEs form a rectangle as depicted in Fig. \ref{fig4_iperez} (a). The reasons for the selection of this geometry will be discussed in the following section. Note that we have the two lengths which are parallel to the $x$-axis and two transversal lengths perpendicular to this same axis.

As in method 1 a clock, to measure the round trip time, is placed at $O$ but now the detector is not fixed. The observer will determine the total time for a series of measurements in which the path length, to be traversed by the PE$_+$, will be increased each time by an amount $x$ while maintaining the returning length $L$ fixed (see Fig. \ref{fig4_iperez} b). In the first measurement the $PE_+$ will travel an initial length $l_0$ whereas the PE$_-$ will travel the returning fixed length $L=l_0+2l_{\bot 0}$ with $L\ge l_n$, where $l_n$ is the maximal length to be traversed by the PE$_+$. In the subsequent measurements the length is increased by a quantity $x=dl_{i+1}=l_{i+1}-l_{i}$ (with $i=0,1,2,\dots, n$), which corresponds to an increment of time $dt_{i+1}=t_{i+1}-t_{i}$. Because of the length $L$ is fixed at all times it must be true that $L=l_{i+1}+2l_{\bot i+1}=l_{i}+2l_{\bot i}$, thus $l_{\bot i+1}=l_{\bot i}-x/2$.

The total time in a measurement is $t_{i}=$time of flight+delay time $=t_{i+} +t_{i-}=l_{i}/u_{+}+L/u_-$. Thus adding the distance $dl_{i+1}$ to the forward path while keeping constant the returning one yields $t_i+dt_{i+1}=(l_i+dl_{i+1})/u_++L/u_-$. Let us subtract $t_i$ and rearrange to get the measured speed: $\bar{u}_{i+1}=dl_{i+1}/dt_{i+1}=u_{+}$. Hence the quantity $dl_{i+1}/dt_{i+1}$ is the slope of the graph for distance versus time, that is, the one-way speed of the PE$_+$. This line of reasoning is again a good procedure for the isotropic system of reference $K$ but for different inertial systems of reference in motion relative to $K$ other values would be obtained.

\subsubsection{Method 2 in $K'$: longitudinal motion}
\label{methodlong}
Consider that the whole instrumentation is moving in the $x$-direction at constant speed $v<c$. Following a similar procedure as in method 1 and keeping in mind our geometry the respective elapsed times for the forward and returning journeys are:
\begin{equation}
\label{eqmet2}
t_{+i}=\frac{l_{i}}{u_+-v}; \qquad t_{-i}=\hat{t}_{-i}+2t_{\bot-i}.
\end{equation}
Here 
\begin{equation}
\label{nose}
\hat{t}_{-i}=\frac{l_i}{u_-+v}
\end{equation}
is the time that the PE$_-$ travels in the direction parallel to the $x$-axis and
\begin{equation}
\label{nose}
t_{\bot-i}=\frac{l_{\bot i}}{u_-}\gamma_{-}
\end{equation}
is the time required in the transversal direction. Taking into account relativistic effects, the observer in $K'$ gets
\begin{equation}
\label{inktime}
t'_{+i}=\frac{l'_{i}}{u_+-v}\gamma^{-2}; \quad \hat{t}'_{-i}=\frac{l'_i}{u_-+v}\gamma^{-2}; \quad t'_{\bot-i}=\frac{l'_{\bot i}}{u_-}\gamma_{-}\gamma^{-1};
\end{equation}
or adding for the total time yields
\begin{equation}
\label{totimek}
t'_i=l'_{i}\gamma^{-2}\biggl(\frac{1}{u_+-v}+\frac{1}{u_-+v}\biggr)+2\frac{l'_{\bot i}\gamma_-\gamma^{-1}}{u_-}.
\end{equation}
It follows that
\begin{equation}
\label{deltatt}
dt'_{i+1}=dl'_{i+1}\gamma^{-1}\biggl[\gamma^{-1}\biggl(\frac{1}{u_+-v}+\frac{1}{u_-+v}\biggr)-\frac{\gamma_-}{u_-}\biggr]
\end{equation}
and
\begin{equation}
\label{uprimepar}
\bar{u}'_{i+1}=\frac{dl'_{i+1}}{dt'_{i+1}}=\frac{\gamma}{\biggl[\gamma^{-1}\biggl(\frac{1}{u_+-v}+\frac{1}{u_-+v}\biggr)-\frac{\gamma_-}{u_-}\biggr]}.
\end{equation}
Note that when $v=0$ the expressions reduce to the case at rest in the isotropic system. And once again we have three unknowns: $u_{\pm}$ and $v$. \textbf{Therefore this method does not provide us the measurement of the one-way speed of any PE in a system in motion relative to $K$}.

\subsubsection{Transversal motion}
 \begin{figure}[htp]
\begin{center}
\includegraphics[width=2.5cm]{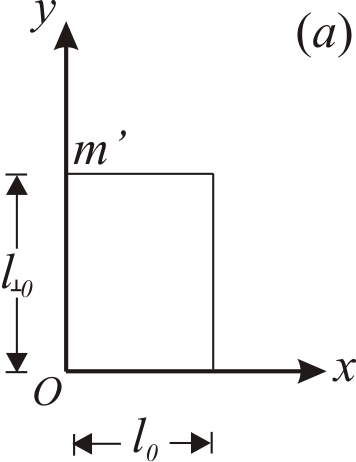}\hspace{1.5cm} \includegraphics[width=2.5cm]{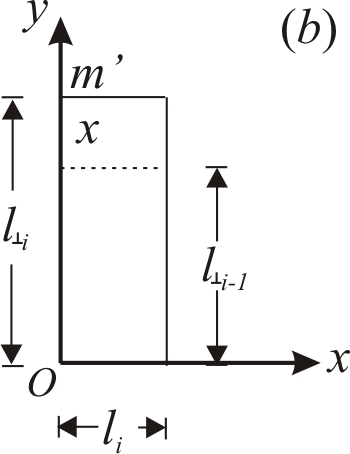}
\caption{Experimental setup rotated $\pi/2$ rad conducted in $K$. ($a$) Initial measurement with $L=l_{\bot0}+2l_0$. ($b$) The rectangle is modified in the following measurements keeping $L$ constant.}
\label{fig6_iperez}
\end{center}
\end{figure}
Now assume first that we are in the isotropic system $K$ and that the experimental setup has been rotated $\pi/2$ rad as shown in the Fig. \ref{fig6_iperez}. Due to the rotation the roles of the lengths have been interchanged hence $l_{\bot i+1}=l_{\bot i}+x$ and $l_{ i+1}=l_{ i}-x/2$ but as long as we stay in this system the calculations turn out to be trivial, they are the same we carried out in subsection \ref{metho2}. 
However, let us analyze the situation in $K'$ which moves relative to $K$. In such case the PE$_+$ must follow a transversal trajectory which can be determined using the Pythagorean theorem as we did before. Hence the times in this direction are
\begin{equation}
\label{transvk}
t_{\bot \pm i}=\frac{l_{\bot i}}{u_{\pm}}\gamma_{\pm}.
\end{equation} 
The returning time is $t_{\bot -i}=\hat{t}_{\bot -i}+t_{-i}+t_{+i}$ where
\begin{equation}
\label{timeskp}
\hat{t}_{\bot-i}=\frac{l_{\bot i}\gamma_-}{u_-}; \qquad t_{\pm i}=\frac{l_{ i}}{u_-\mp v}.
\end{equation}
Hence, considering relativistic effects the total time in $K'$ becomes
\begin{equation}
\label{ttotalk}
t'_{\bot i}=l'_{\bot i}\gamma^{-1}\biggl(\frac{\gamma_+}{u_+}+\frac{\gamma_-}{u_-}\biggr)+\frac{2l'_{i}\gamma^{-2}\gamma^2_-}{u_-}.
\end{equation}
Hence
\begin{equation}
\label{ }
dt'_{\bot i+1}=dl'_{\bot i+1}\gamma^{-1}\biggl(\frac{\gamma_+}{u_+}+\frac{\gamma_-}{u_-}-\frac{\gamma^{-1}\gamma^2_-}{u_-} \biggr),
\end{equation}
and the measured speed is
\begin{equation}
\label{sppedutrans}
\bar{u}'_{\bot i+1}=\frac{dl'_{\bot i+1}}{dt'_{\bot i+1}}=\frac{\gamma}{\biggl(\frac{\gamma_+}{u_+}+\frac{\gamma_-}{u_-}-\frac{\gamma^{-1}\gamma^2_-}{u_-} \biggr)}
\end{equation}
Again we still have the unknowns $u_{\pm}$ and $v$. Note also that when $v=0$ the expression reduces to the case at rest.

\subsubsection{Determination of the speeds}

To determine the unknown speeds we use the same trick as we did before. We make $u\equiv u_-\approx u_+ $ and thus Eqs. \eqref{uprimepar} and \eqref{sppedutrans} become
\begin{equation}
\label{uuprime}
\bar{u}'=\frac{u\gamma \gamma^{-1}_{u}}{(2\gamma^{-1}\gamma_u-1)} \quad \textrm{and}  \quad \bar{u}'_{\bot }=\frac{u\gamma\gamma^{-1}_u}{(2-\gamma^{-1}\gamma_u)},
\end{equation}
respectively.  Also from the previous expressions one can solve for $u$ and $v$ in terms of the measured speeds. That is
\begin{equation}
\label{vvv}
v=uc\Biggl[\frac{3(b^2-1)}{c^2(1+2b)^2-u^2(2+b)^2}\Biggr]^{1/2}, 
\end{equation}
where
\begin{equation}
\label{wherevvv}
b=\frac{\bar{u}_{\bot}'}{\bar{u}'}; \qquad  u=\frac{3\bar{u}_{\bot}' \bar{u}'(\bar{u}'+2\bar{u}_{\bot}')}{\bar{u}_{\bot}'^2+4\bar{u}_{\bot}'\bar{u}'+4\bar{u}'^2}.
\end{equation}
If we assume that $u_+= u_- = c$ from Eqs. \eqref{uuprime} we would expect that $\bar{u}'=\bar{u}'_{\bot } = c$ which also justifies why measurements of the speed of light are consistent among different methodologies. This case can be carried out if we replace the cables for mirrors forming our rectangular setup \cite{michelson3}.

\section{Discussion}

In the previous subsection we have selected a rectangular geometry for our setup which is clearly a closed path constituted by the forward path and the returning path. The selection of the geometry is arbitrary but we use the rectangular one just to simplify the calculations and illustrate our points. Other geometries will lead to more complex calculations. 

What is worth noticing from section \ref{metho2} is that most physicists naively assume that the length $L$ of the returning path (usually determined by the length of a cable) is constant. They consider that the returning speed $u_-$ of the electric field is constant and therefore conclude that the delay time $t_-=L/u_-$ too. This is true as long as the cable is at rest in $K$. But when the cable is in motion we have seen above that there is a section of the cable that must move parallel to the motion, while the remaining one must move perpendicularly, therefore the sections of the cable that moves perpendicular to the motion do not undergo length contraction. Their belief is justified since the measuring rods in the system $K'$ suffer the same length contraction as the cables and therefore an observer in this system would not realize that the returning path has been affected by the motion; he would obtain the same value for the returning path no matter if he were at rest or in relative motion (invariance). This fact may suggest that this method provides the measurement of the one-way speed of light in any inertial system. However, we have readily shown that what the observer in motion is really measuring is a speed which depends on the speeds $u_{\pm}$, $v$ and $c$ and therefore the one-way speed of any PE cannot be determined by these methods unless the measurements were conducted at rest in the isotropic system. And since we do not know whether the earth is the isotropic system (most probably it is not) their claim that the reported value corresponds to the one-way speed of light losses its validity.

As an illustration of the consistency of our results, let us make a comparison with those found by Greaves et al. \cite{greaves}. First, they assumed that the earth is an isotropic system of reference and used a long cable in which they considered that the returning information travels at the speed $u_-=2c/3\approx 2\times 10^8$ m/s \cite{greaves3}. Then they reported two measurements of the speed of light as $\bar{c}_1=3.016\times 10^8\;\pm 0.071\times 10^8$ m/s and $\bar{c}_2=3.004\times 10^8\;\pm 0.085\times 10^8$ m/s. On the contrary, based on the work of Mansouri and Sexl \cite{mansouri}, we assume the system $K$ to be attached to the cosmic microwave background radiation and the system $K'$ attached to the earth which is relatively moving at speed $v\approx 5 \times 10^5 \, \pm 10^5$ m/s. If we assume their value of $u_-$, and $u_+=c$, the measured speeds of light predicted by Eqs. \eqref{uprimepar} and \eqref{sppedutrans} yield $\bar{u}'= 300 \,433 \,311$ m/s and $\bar{u}'_{\bot}= 299 \, 793\, 401$ m/s, which within the uncertainty are in agreement with those of Greaves et al. This suggest itself that their reported values, in fact, correspond to a two-way measurement. 

The appropriate experimental technique to measure the one-way speed in the system $K'$ is by the use of two synchronized clocks placed at the endpoints, but, an accurate synchronization process requires the knowledge of the one-way speed of light $c'_{\pm}$ which causes a circular reasoning. This problem and other synchronization methods can be found elsewhere \cite{zhang,guerra1,guerra2,abreu,iyer1,feenberg,mansouri}.

Finally, it is important to mention that interferometric experiments like the Michelson-Morley experiment \cite{michelson1,muller,muller1,wolf,perez1} are also two-way experiments and are readily explained by method 1; we just have to set $u=c$ in the corresponding equations. Iyer and Prabhu have derived a more general expression for the one-way speed of light in $K'$, they found the relation $c'=c^2(c\pm\mathbf{\hat{k}'}\cdot \mathbf{v})^{-1}$, where $ \mathbf{\hat{k}'}$ is a unit vector that points in the direction of energy flow of the light beam as determined in $K'$, $\mathbf{v}$ is the velocity vector of $K'$ relative to $K$ and $c$ is two-way speed of light. Here once again $\mathbf{v}$ is the velocity as measured in the isotropic system. The reader can easily verify from this equation that the harmonic mean of the speed of light, in any direction, is $c$. This fact explains the negative outcome of the experiment and, therefore, favors the existence of the isotropic frame.

\section{Concluding Remarks} 

We have analyzed in detail two common methods for the determination of the speed of any PE. One of the main goals of this contribution was simply to elucidate the physics behind the measuring processes. Our analysis was based on the belief that there exists at least one isotropic system where ME is valid; if this is true our results show that the one-way and the two-way measured speed of any PE for inertial systems in motion relative to the isotropic system are in fact function of the speed of the system in motion. However, since in most experiments the paths followed by the PEs are closed, one is really impeded to determine its one-way value. In this sense we have justified with some examples why the two methods, independently of their spatial orientations, yield the measured speed equal to $c$ provided that $u_+\approx u_-\approx c$.

Finally, our study suggests a profound analysis of the experimental methods in which closed paths are involved. We have seen that the notion of closed path is not to be restricted to light but also to any kind of PE. In this respect, the present investigation was intended to boost and encourage the experimental and theoretical investigations to overcome these technical conundrums. 

\subsection*{Acknowledgments}

I am grateful to Georgina Carrillo for helpful comments.

\end{document}